\documentclass[preprint2]{aastex}

\usepackage{aastex_plus}
\shorttitle{Evidence for Steady Heating in Moss}                 
\shortauthors{Brooks}
\begin{document}

\title{Flows and Motions in Moss in the Core of a Flaring Active Region: Evidence for Steady Heating.}
\author{David H. Brooks \altaffilmark{1,2} and Harry P. Warren}
\affil{Space Science Division, Naval Research Laboratory, Washington, DC 20375}
\altaffiltext{1}{George Mason University, 4400 University Drive, Fairfax, VA 22020}                                  
\altaffiltext{2}{Present address: Hinode Team, ISAS/JAXA, 3-1-1 Yoshinodai, Sagamihara, Kanagawa 229-8510, Japan}
\email{dhbrooks@ssd5.nrl.navy.mil}

\begin{abstract}
We present new measurements of the time variability of intensity, Doppler and non-thermal velocities
in moss in an active region core 
observed by the EUV Imaging Spectrometer 
on Hinode in 2007, June. The measurements are derived from spectral profiles of the \ion{Fe}{12} 195\,\AA\, line. 
Using the 2$''$ slit, we 
repeatedly scanned 150$''$ by 150$''$ in a few mins.  
This is the first time it has been possible to make such velocity measurements in the moss,
and the data presented are the highest cadence spatially resolved maps of moss Doppler and non-thermal velocities
ever obtained in the corona. 
The observed active region produced numerous C- and M- class flares with several occurring in the core close to
the moss. The magnetic field was therefore clearly changing in the active region core, so we 
ought to be able to detect dynamic signatures in the moss
if they exist. Our measurements of moss intensities agree with previous studies in that a less 
than 15\% variability is seen over a period of 16 hours. 
Our new measurements
of Doppler and non-thermal velocities reveal no strong flows or motions in the moss, nor 
any significant variability in these quantities. 
The results confirm that moss at the bases of 
high temperature coronal loops is heated quasi-steadily. They also show that quasi-steady heating can contribute significantly
even in the core of a flare productive active region. Such heating may be impulsive
at high frequency, but if so it does not give rise to large flows or motions.
\end{abstract}

\section{Introduction}
One of the most important unsolved problems in astrophysics is the question of how the solar corona 
is heated. 
Directly related to this problem is understanding the evolution of
the emission in active regions, particularly in the core. 
Recent observations from {\it Hinode}
\citep{kosugi_etal2007} are providing a comprehensive view of active regions. They appear to be composed of at least two
dominant loop populations: core loops that are seen evolving at multi-temperatures, and
peripheral cool loops that develop downflows  
\citep{ugarteurra_etal2009}. The former 
appear to be rooted in unipolar magnetic concentrations \citep{brooks_etal2008}.
The latter
are seen extending to great heights and appear to be associated with coronal rain. 
Fan structures around the edges of active regions show strong outflows \citep{sakao_etal2007}
that may be significant sources of the slow solar wind (\citealt{doschek_etal2008,harra_etal2008}).

A key question is what is the time-scale of heating in the cores of active regions? 
Is steady heating important, or is everything dynamic? If the heating is impulsive, is the repetition
time between events shorter than a coronal loop cooling time so that the impulsive heating is effectively steady?
Hydrostatic
models are able to reproduce the emission from high temperature loops seen in active regions,
but they have difficulty reproducing the emission from lower temperature (warm) overdense loops 
\citep{warren&winebarger_2006,klimchuk_2006}.
Hydrodynamic simulations of active regions \citep{warren&winebarger_2007}
or modeling based on low frequency impulsive heating by nanoflares \citep{patsourakos&klimchuk_2008}, 
improve the agreement with the warm loop observations, but discrepancies still remain.
It is not always clear whether these discrepancies are
related to the hydrostatic or hydrodynamic modeling, or to the extrapolation and
modeling of the morphology of the magnetic
field \citep{brooks&warren_2008}. 

One region where the heating could be quasi-steady is in transition region moss.
Moss is
the reticulated pattern of emission in active regions that typically evolves slowly over long
time-scales and shows only low-level variability \citep{berger_etal1999}. Indeed hydrostatic modeling of moss is
able to reproduce the observed intensities well - to within a filling factor 
\citep{warren_etal2008b}.
Though the general pattern evolves slowly, fine-scale dynamics are observed on a local scale,
and this has been interpreted as variability in the overlying spicular material rather than in 
the moss itself \citep{depontieu_etal1999}. This picture of fine-scale dynamics within a
slowly evolving large-scale pattern also well describes the unipolar regions
at the footpoints of hot loops \citep{brooks_etal2008}. The fact that moss
is thought to be the emission from the footpoints of hot loops \citep{tberger_etal1999,martens_etal2000},
suggests that there may be a close relationship
between these patterns and we are investigating this issue. \citet{tripathi_etal2008}
recently presented an example where the moss 
maps the magnetic flux closely, though \citet{depontieu_etal2003} did not find any spatial correlation
with the photospheric magnetic field.

It has been argued that moss in highly evolved active regions may be the only regions in the corona
where the heating could be steady because these are the only places where the magnetic field is
not changing rapidly \citep{antiochos_etal2003}. Detection of flows and motions in the
moss, however, would reveal dynamic signatures for the first time. This letter presents
an analysis of the time-variability
of flows and motions in moss over 16 hours of EIS observations of the 
core of a flaring active region
observed in 2007, June. We show that the Doppler and non-thermal velocities are no larger than typical
quiet Sun values in
the moss, and that they do not vary substantially over the observations period.
These results support the view that moss at the bases of hot loops is heated quasi-steadily.

\begin{figure*}
\centering
\includegraphics[width=0.32\linewidth]{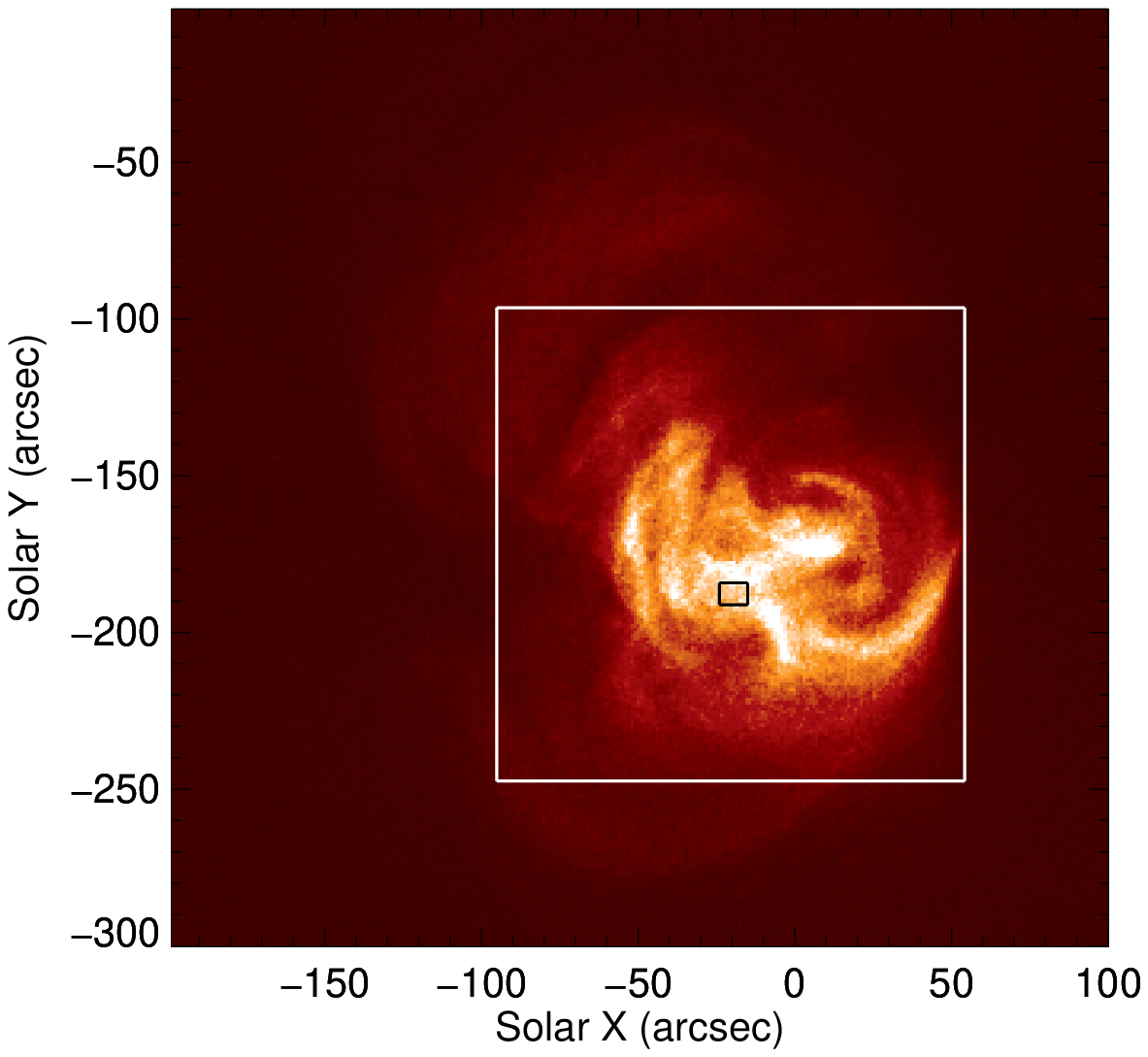}
\includegraphics[width=0.32\linewidth]{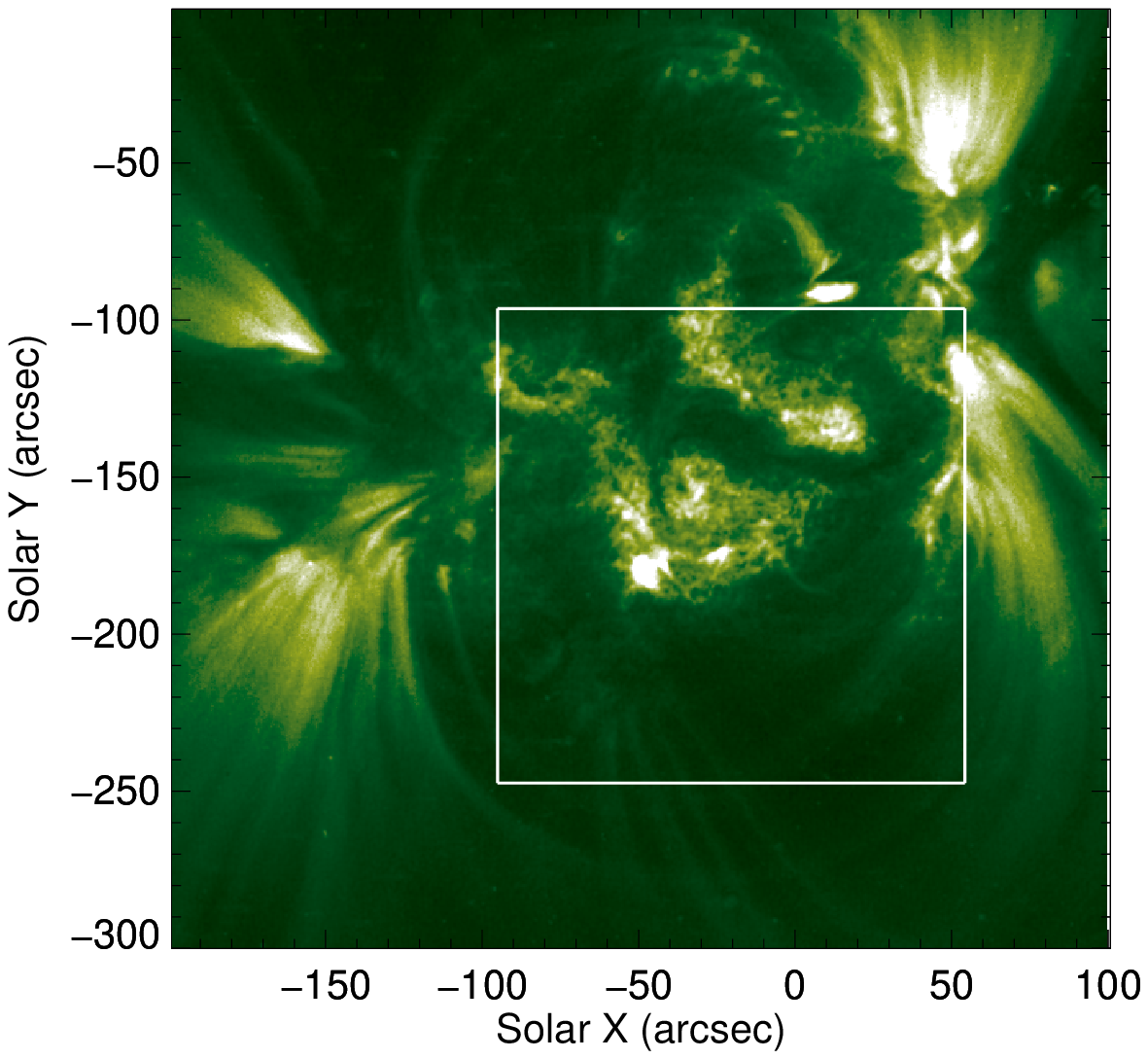}
\includegraphics[width=0.32\linewidth]{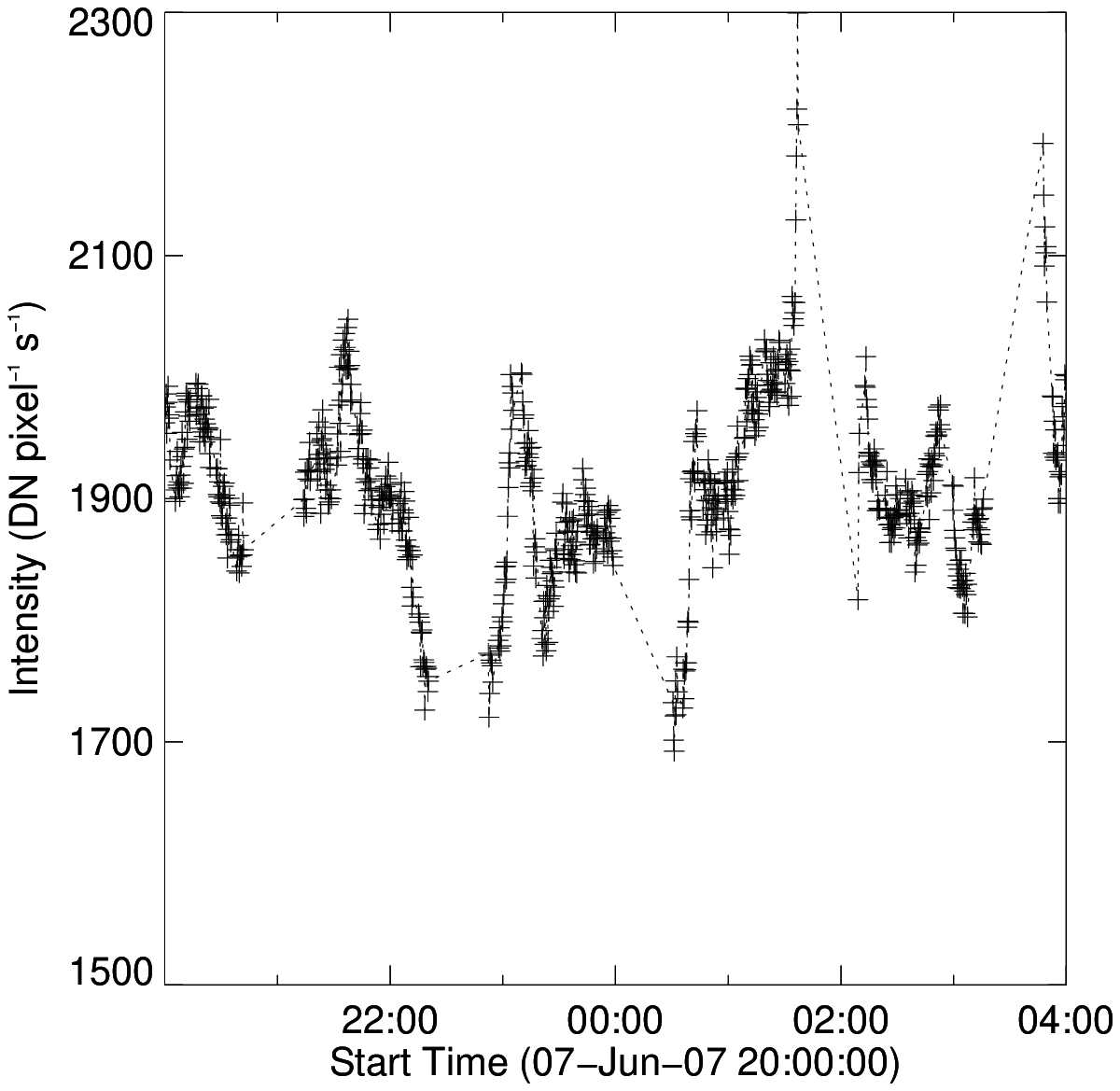}
\caption{Left panel: XRT Open/Ti Poly image taken at 18:19:36UT on 2007, June 7. 
Center panel: TRACE 171\,\AA\, filter image taken at 18:19:05UT on 2007, June 7.
Right panel: XRT light curve for the small black box in the left panel.
The EIS FOV is overlaid on both the XRT and TRACE images as a white box. 
XRT and TRACE movies of the region are
available in the electronic version of the manuscript (video\_fig1a and video\_fig1b).}
\label{fig1}
\end{figure*}

\begin{figure}[ht]
\centering
\includegraphics[width=0.7\linewidth]{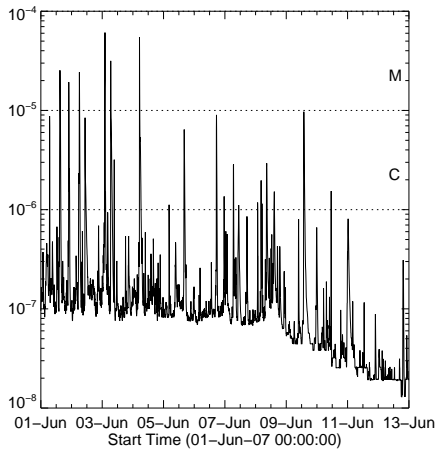}
\caption{GOES-11, 5 min X-ray flux (1--8\,\AA). }
\label{fig2}
\end{figure}

\section{Observations and Data Reduction}
The EIS instrument obtains high resolution spectra in the wavelength ranges 171--212\,\AA\,
and 245-291\,\AA\, and has a spectral resolution of 0.0223\,\AA\, and 1$''$
spatial pixels \citep{culhane_etal2007}. Typical exposure times for active regions 
are 15--30s so that it takes tens of mins to several hours for the EIS instrument to step 
its slit over 
the full extent of a large active region. Thus observational cadence or spatial coverage are usually traded off for    
diagnostic information when making observations with a spectrometer in comparison with using a broad-band filter 
imager such as TRACE \citep[][Transition Region
and Coronal Explorer]{handy_etal1999} or {\it Hinode}/XRT \citep[][X-ray Telescope]{golub_etal2007}.
The EIS effective area peaks near the strong 
\ion{Fe}{12} 195.119\,\AA\, emission line, however, so that very short exposures (1s) can still yield
good counts in this line on a relatively bright target. It is therefore possible to make rapid scans
over a wide area in strong lines only, effectively trading off wavelength coverage to retain high 
cadence and a large FOV.
We use data obtained in this observing mode in this letter.

\begin{figure*}
\centering
\includegraphics[width=0.9\linewidth]{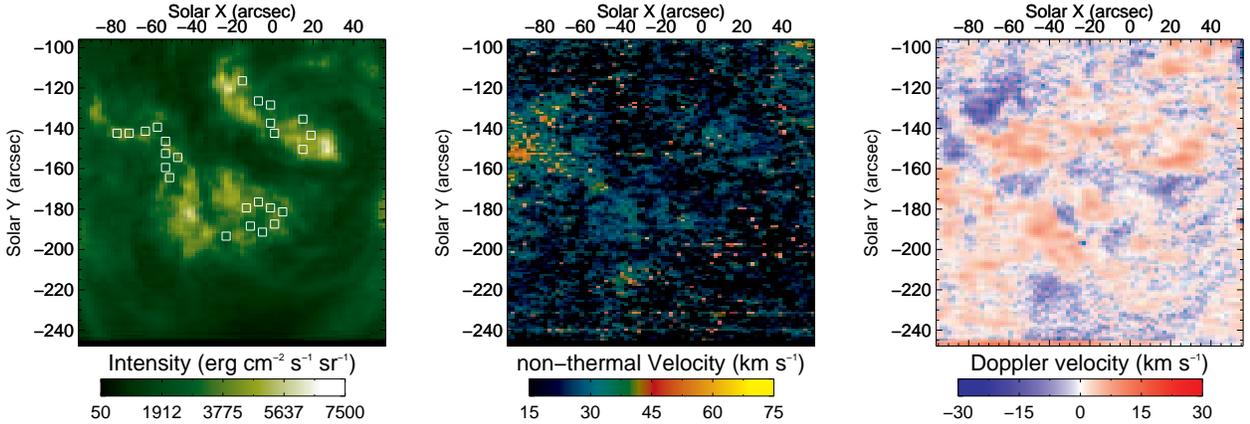}
\caption{EIS measurements derived from the \ion{Fe}{12} 195.119\,\AA\, 
line. The images are scaled linearly in the ranges
of the color bars. A movie of the complete observing period is available in the electronic
edition (video\_fig3). Residual orbital velocity effects have
been removed from the movie for clarity. The boxes on the intensity image show the
25 locations analyzed in Figure \ref{fig4}. Some images show artifacts at the bottom because
they were shifted during coalignment.}
\label{fig3}
\end{figure*}

{\it Hinode} tracked NOAA active region 10960 from limb to limb in June 2007. EIS repeatedly
ran an observing sequence CHROMO\_EVAP\_RASTER which takes 1s exposures with the 2$''$ slit and scans
an area of 150$''$ by 150$''$ in a few mins. Figure \ref{fig1} shows XRT and TRACE images
of AR 10960 with the EIS FOV overlaid. The movie associated with Figure \ref{fig1} 
shows that AR 10960 evolved slowly in XRT images, 
and this is confirmed by the X-ray light curve shown in Figure \ref{fig1}. The intensity
observed by XRT in a boxed region above the moss varies by less than 15\% over the 8 hour period
shown. The TRACE movie 
shows that the pattern of moss in the core of the region is also 
remarkably stable. 
Nevertheless, this region was flare productive: 
Figure \ref{fig2} is a GOES plot of the X-ray activity, which shows numerous C- and M- class flares occuring
during its passage across the disk, and also a gradual
decline in activity.
These characteristics make AR 10960
an ideal candidate for studying whether impulsive or steady heating dominates in the core of
the region. 

The EIS data analyzed here were obtained between 18UT on 2007, June 7, and 10UT on 2007, June 8.
Several C- class flares did occur during these 16 hours though 
not all were cleanly observed by
{\it Hinode} or TRACE because of South Atlantic Anomaly (SAA) and night-time passage. 
At least two flares are seen, however, in our XRT and TRACE movies
and it is clear that they
occurred in the core of the region near the neutral line with the potential to disrupt
or interact with the moss. 

The EIS data were converted
to physical units and processed 
using the default options for the SolarSoft routine EIS\_PREP.
The thermal orbital variation in line centroid
position was estimated from single Gaussian fits to the 
\ion{Fe}{12} 195.119\,\AA\, line profiles 
at every pixel of every raster scan. This effect and the spectral line tilt
were removed from the data prior to further analysis.
The zero-point velocity for this operation was set by measuring the wavelength of the \ion{Fe}{12} 
line averaged over the complete dataset. 
The {\it Hinode} pointing has an orbital variation on the order of 1--2 pixels in
the X- and Y- directions \citep{shimizu_etal2007}. This effect was removed by coaligning
the single Gaussian fit images by cross-correlation.

Since the 2$''$ slit instrumental width was not measured in the laboratory
pre-launch, we estimated it using measurements of the \ion{Fe}{10} 184.536\,\AA\, line from 
a large raster scan of the quiet Sun taken on 2007, November 6. 
From these data, we found that the 2$''$ slit instrumental width was 
approximately 19\% wider than that of the 1$''$ slit. Taking the on-orbit
value for the 1$''$ slit from 
\citet{brown_etal2008} we estimated the
2$''$ slit instrumental width to be 0.066\,\AA.
The non-thermal velocities 
were calculated from the line widths assuming this measured instrumental width
and the appropriate thermal width at a temperature of 1.35MK. 

The \ion{Fe}{12} 195.119\,\AA\, line is blended with a very weak density sensitive line at 195.179\,\AA.
and we found that single Gaussian fits to our data 
overestimated the Doppler and non-thermal velocities as a result of density variations. 
Our measurements are therefore
derived from double Gaussian fits where the separation of the line centroids
is fixed at 0.06\,\AA\, and the widths of the two components are forced to be the same.
These fit parameter constraints were suggested by
\citet{young_etal2009} and we found that their method
worked well on our data, even in single pixels.

\section{Results}
Figure \ref{fig3} shows an example of the intensity, non-thermal, and Doppler velocity maps
obtained from two component Gaussian fits to the EIS data. 
The moss is only weakly visible in the non-thermal velocity maps, and little variation in
the patterns is seen over 16 hours of data (see the movie 
associated with Figure \ref{fig3}). 

We analyzed the time variability of these quantities in 25 locations in the moss by
visually selecting pixels that are not affected (in the line of sight) by any
flaring activity, and do not lie close to the edges of the moss. These edges border
dark areas that have been shown to have enhanced non-thermal velocities and outflows
in other active regions \citep{doschek_etal2008}. Larger blue-shifts in these areas
can also be seen in Figure \ref{fig3}.

Our aim was to probe the properties of the moss on smaller spatial
scales than \citet{antiochos_etal2003}-- they used a minimum box size of 10$''$$\times$10$''$-- and to reduce smoothing of the measurements that
results from averaging over large regions.
Variations in 
intensities in single pixels can result from not sampling 
precisely the same spatial location in consecutive rasters because of the {\it Hinode}
orbital pointing variation, however. Furthermore, increased
signal in the box improves the reliability of measurements of the contribution from the blend. Therefore, the 
box size was chosen to be $\pm$ 2 pixels around the selected pixels as a compromise.
The velocities here were calibrated to the zero point velocity
by comparing the time-averaged wavelength from single and double Gaussian fits to the data at 
each location. 
Nevertheless, the absolute values are not well constrained. 

\begin{figure*}
\centerline{
\includegraphics[width=0.30\linewidth,clip]{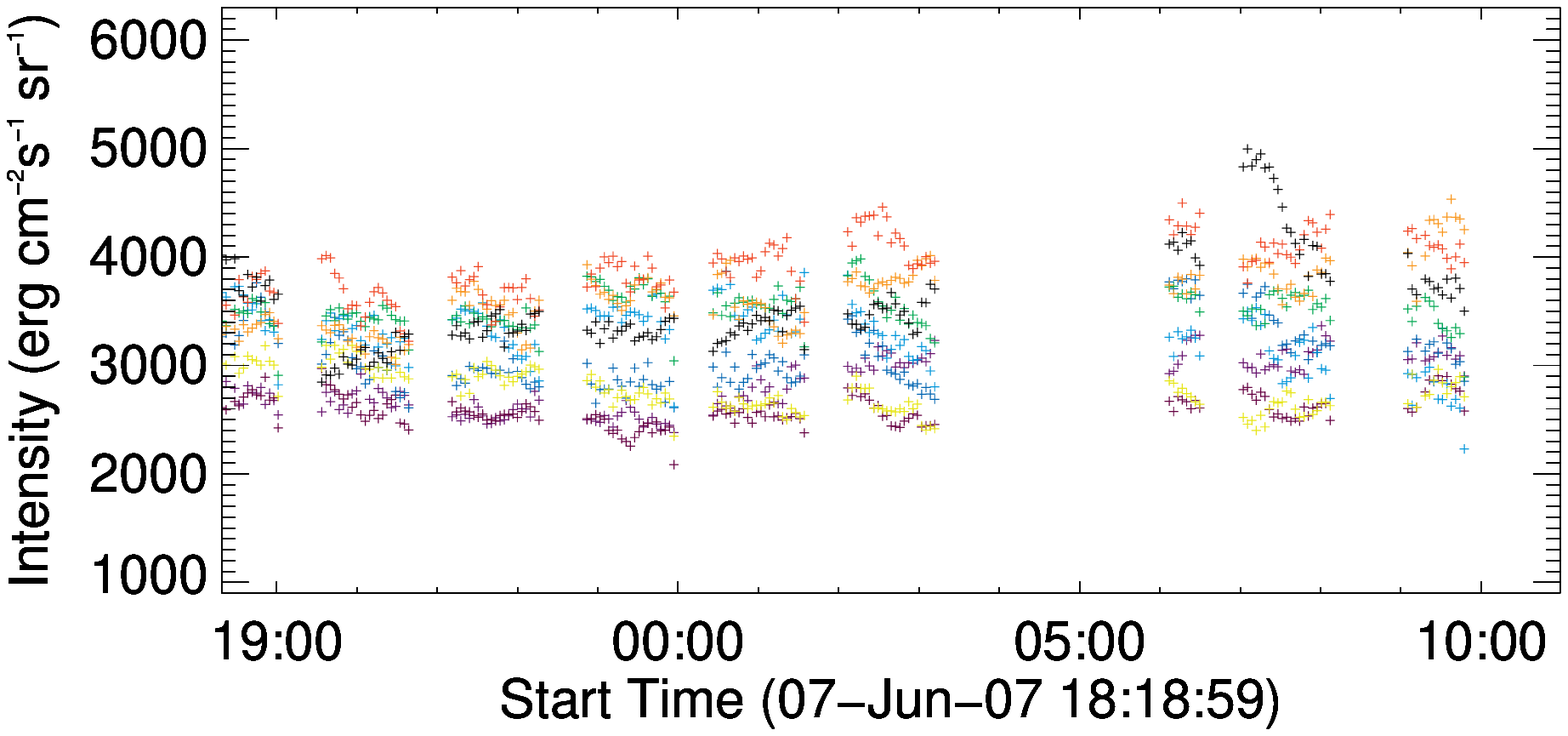}
\includegraphics[width=0.30\linewidth,clip]{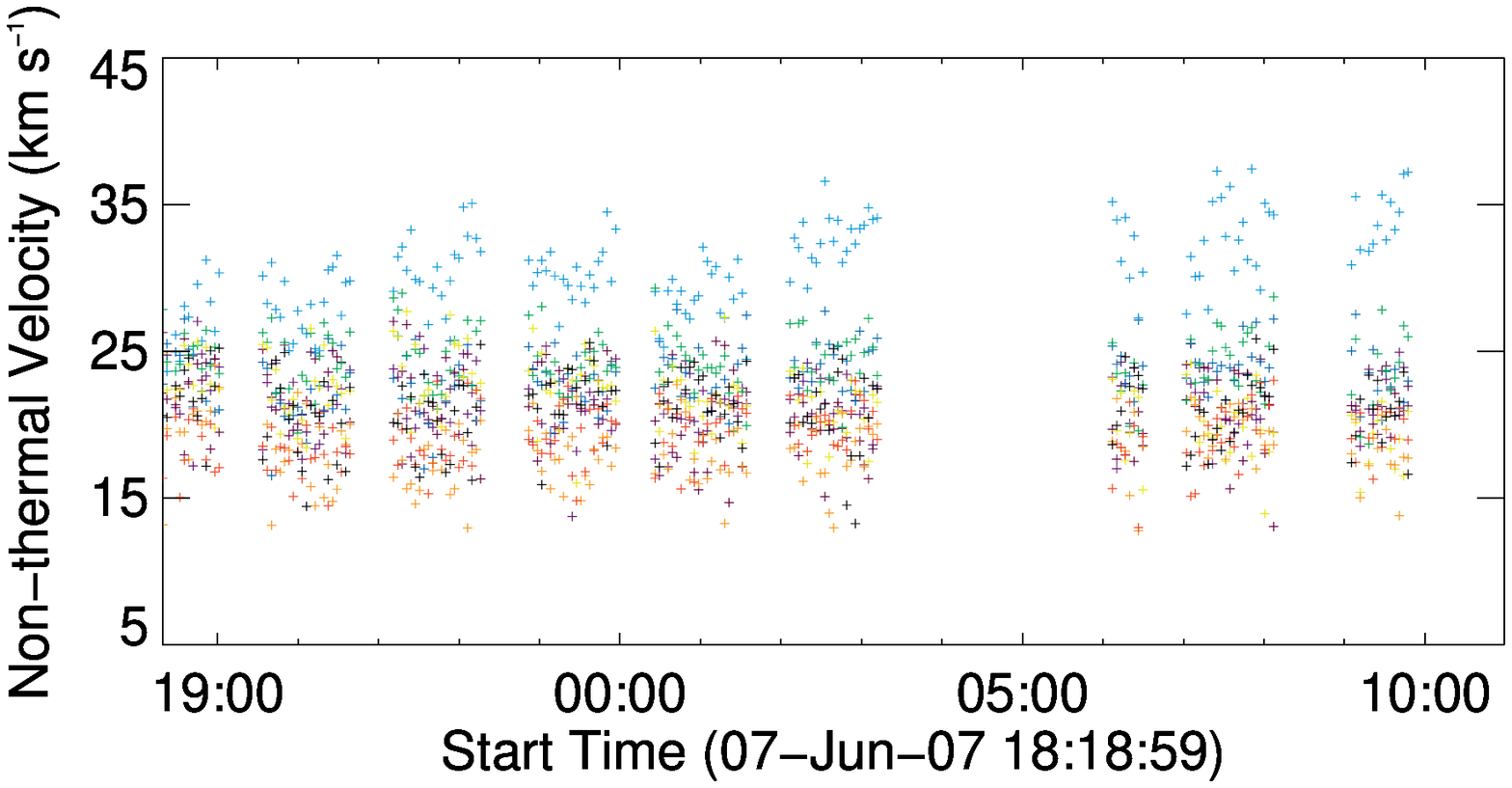}
\includegraphics[width=0.30\linewidth,clip]{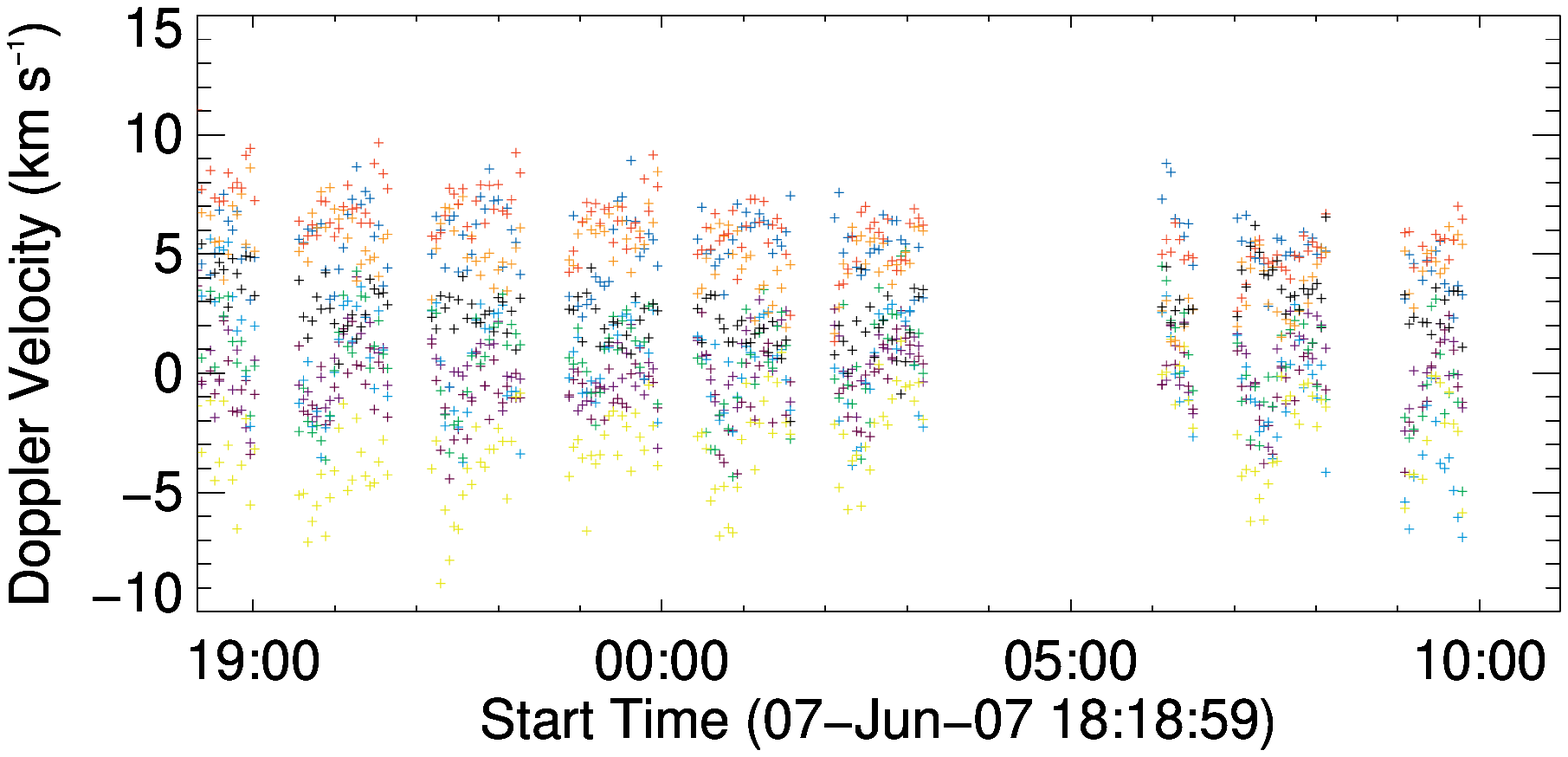}}
\centerline{
\includegraphics[width=0.30\linewidth,clip]{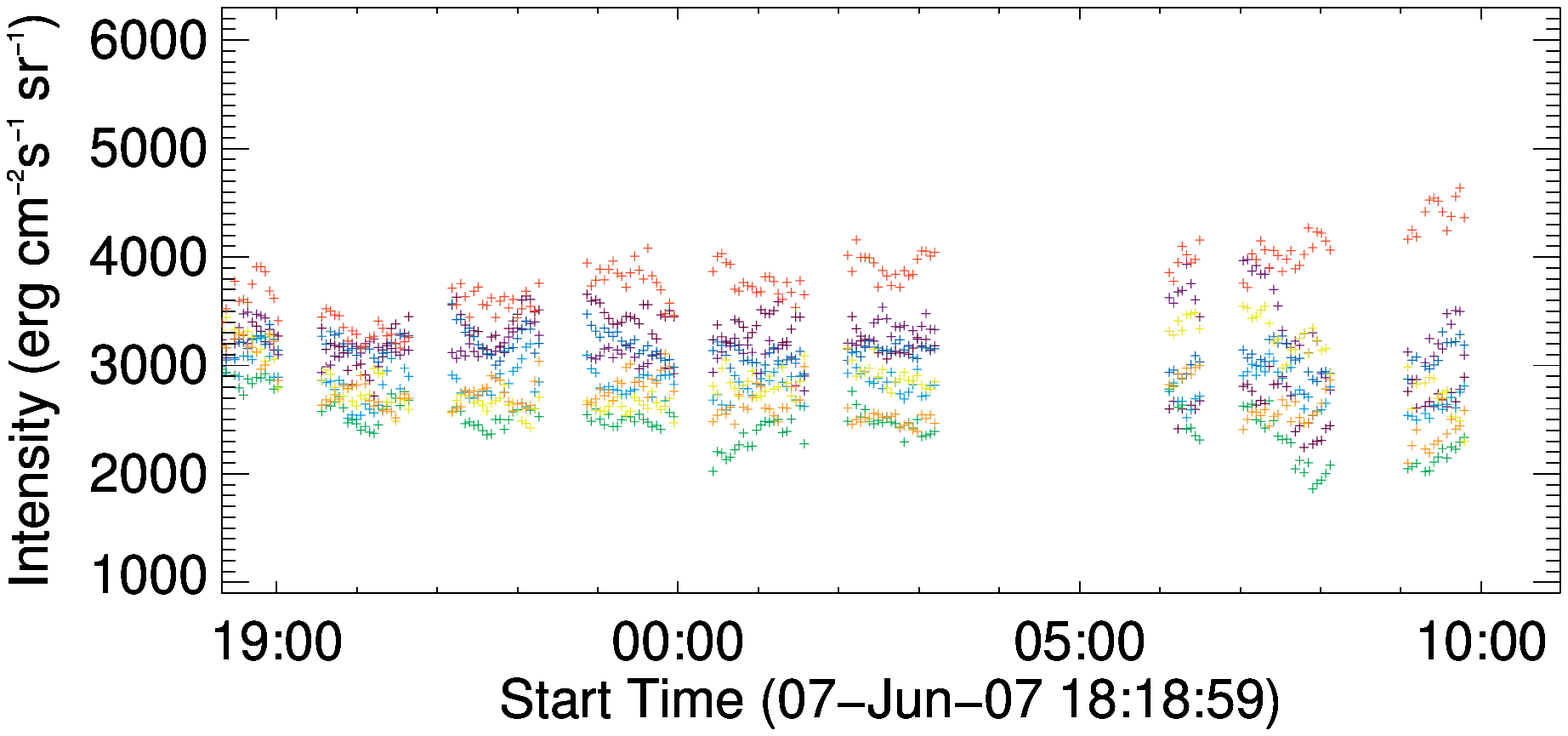}
\includegraphics[width=0.30\linewidth,clip]{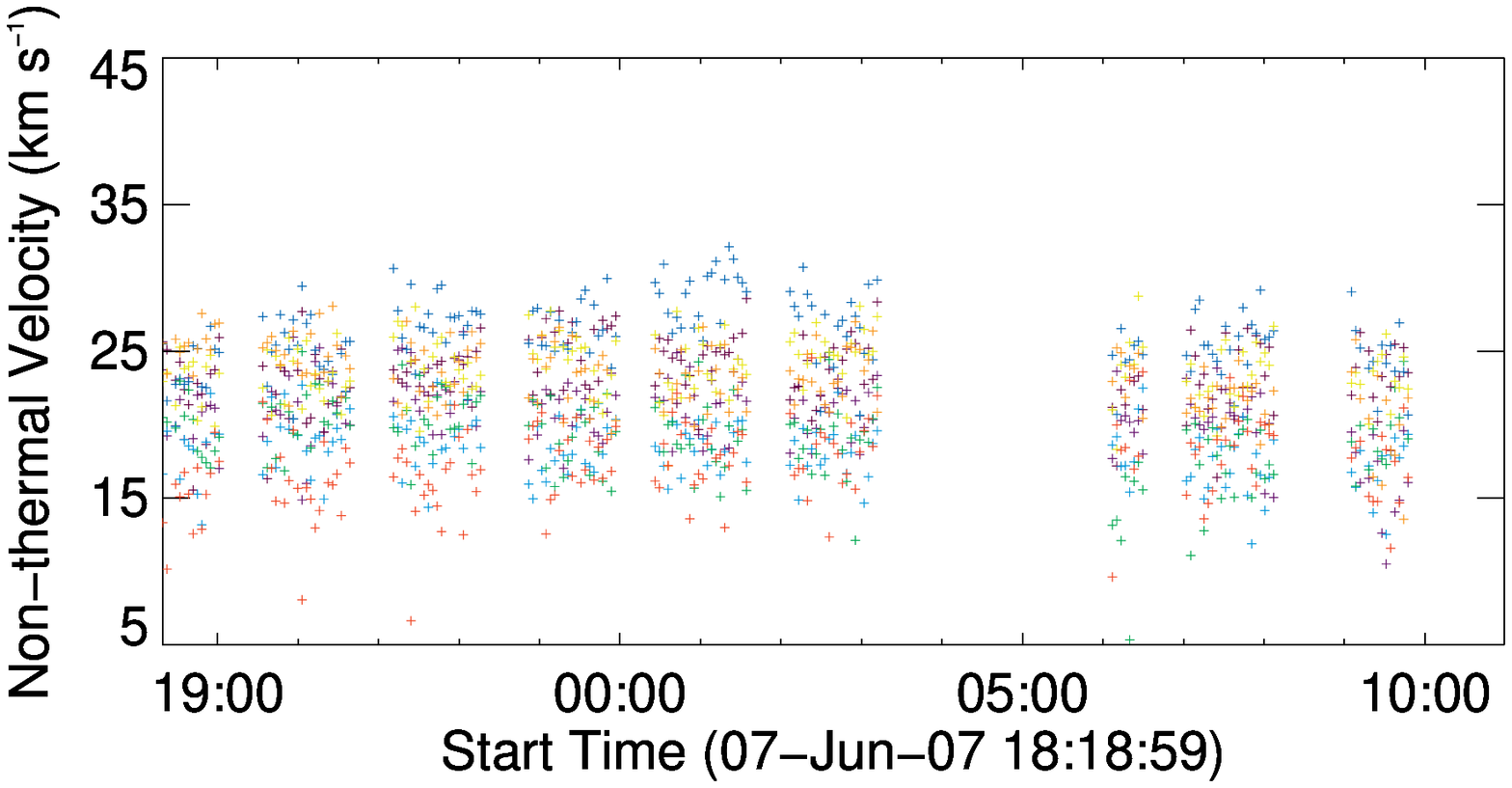}
\includegraphics[width=0.30\linewidth,clip]{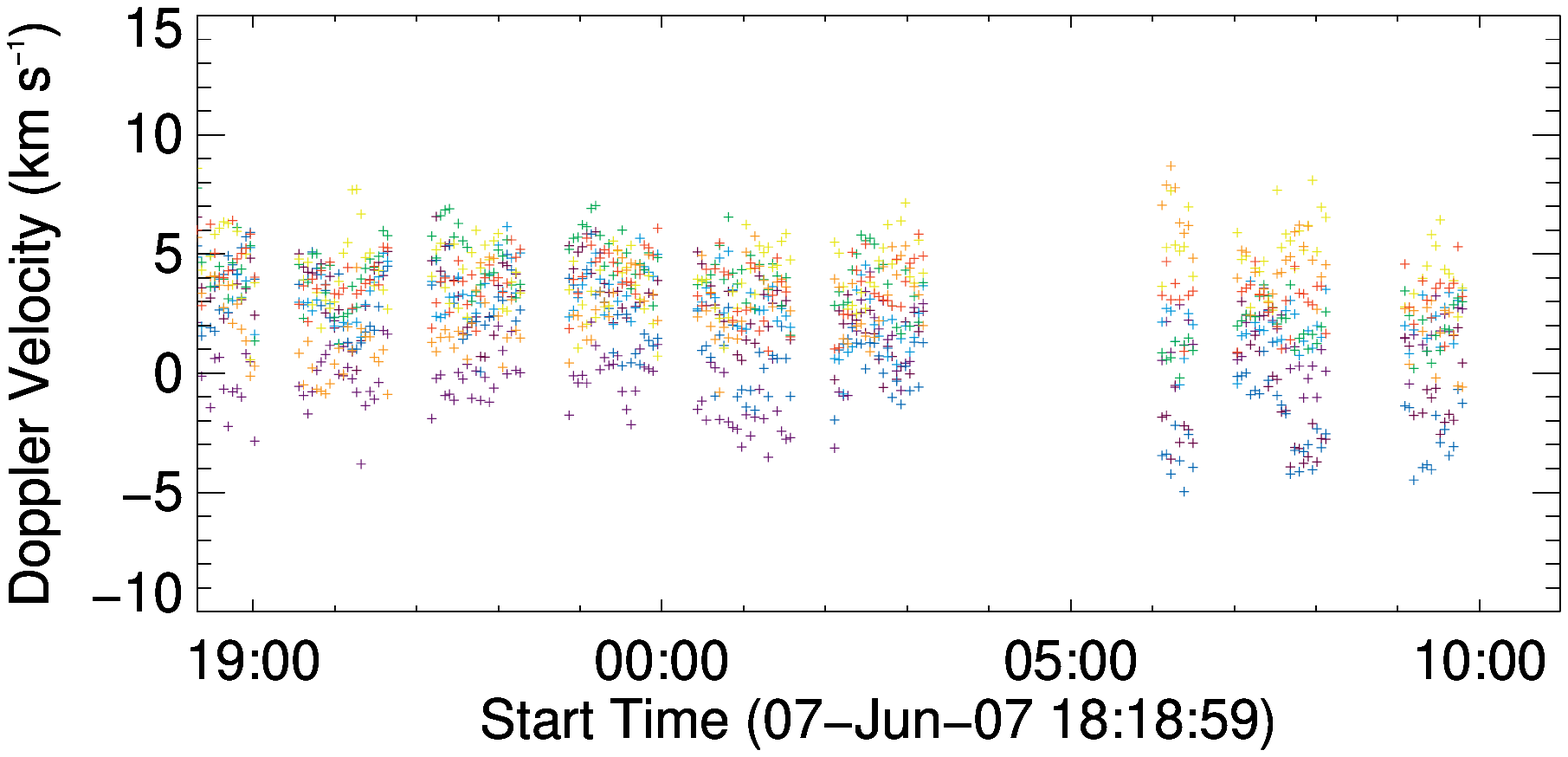}}
\centerline{
\includegraphics[width=0.30\linewidth,clip]{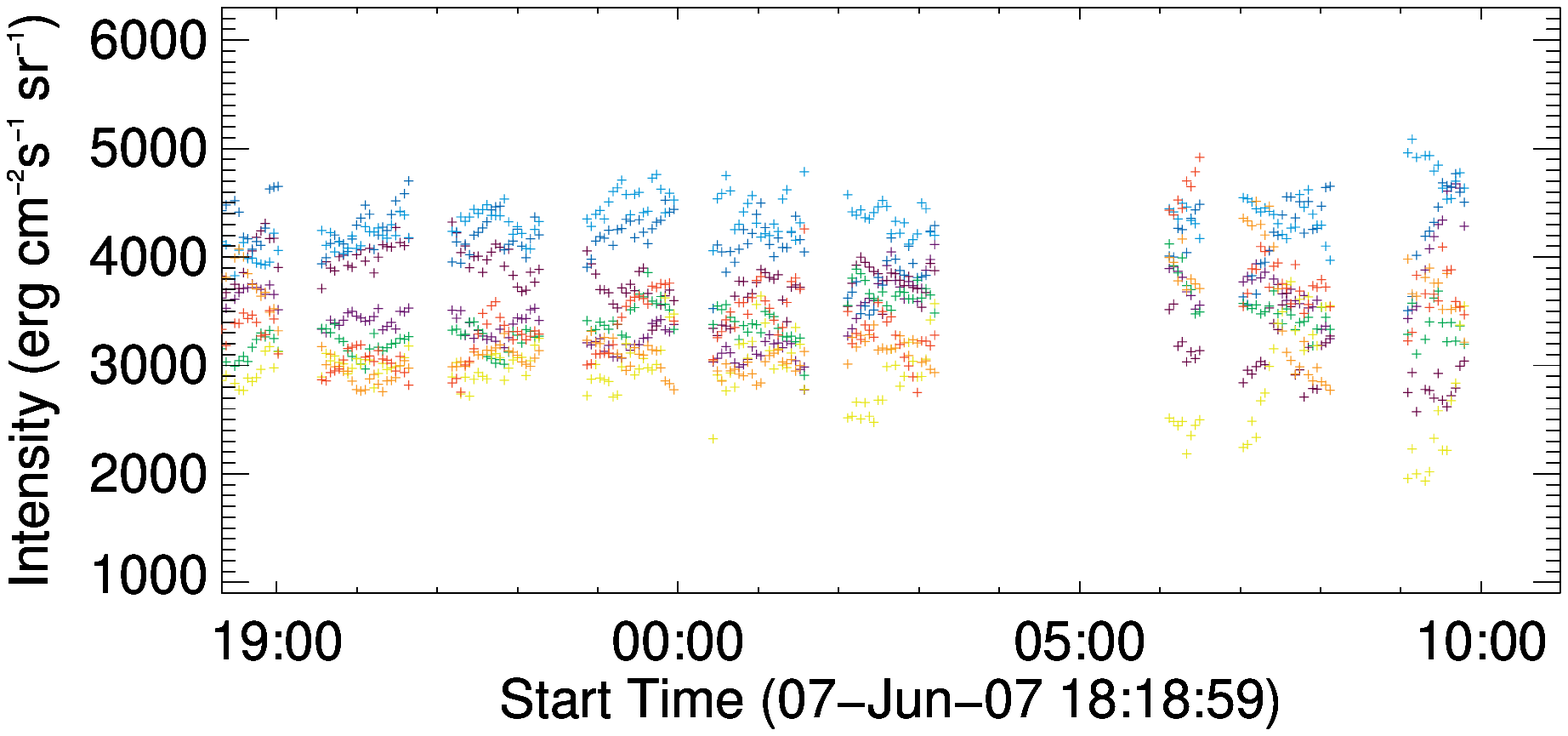}
\includegraphics[width=0.30\linewidth,clip]{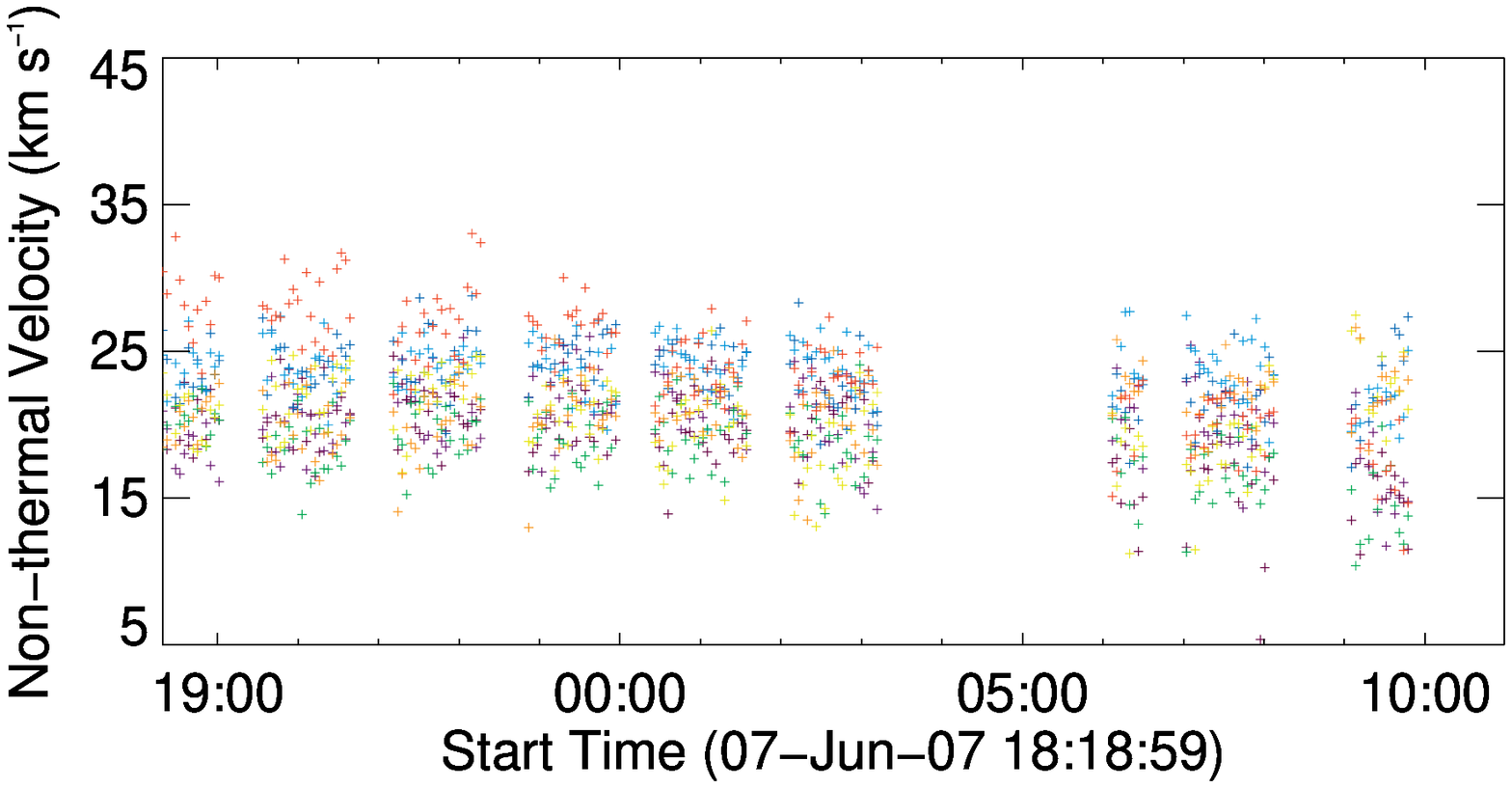}
\includegraphics[width=0.30\linewidth,clip]{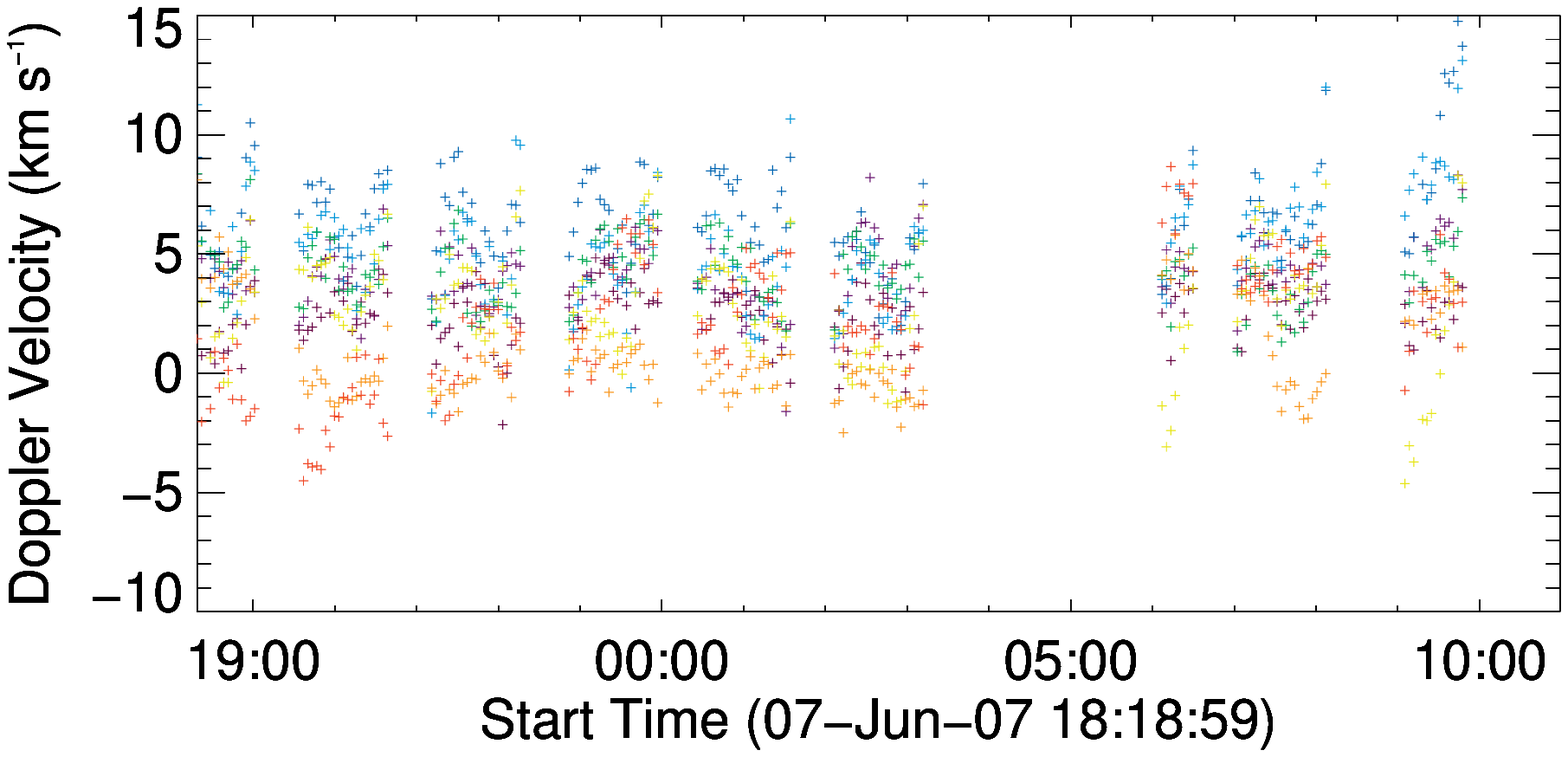}}
\caption{Variation of intensities (1st column), non-thermal velocities (2nd column), and 
Doppler velocities (3rd column) as a function of time in the 25 locations 
identified in Figure \ref{fig3}. For clarity of presentation, each panel of each column 
shows separate colored curves for only 8 or 9 locations. 
}
\label{fig4}
\end{figure*}

The results for the full 16 hour time-period are shown in Figure \ref{fig4}. 
Residual effects from, e.g., cosmic ray hits that were not completely removed,
have been ignored by only quoting results from the 98\% of significant values.
Consistent with previous results obtained by TRACE, the
high spectral resolution EIS obervations show that the \ion{Fe}{12} 195.119\,\AA\,
intensities do not vary significantly over time. The standard deviation of the intensities
from the 25 locations over 16 hours is less than 15\% of the average intensity in
each region. This is less than the variation of the average intensity
from location to location (30-35\%). 

Our new measurements of Doppler and non-thermal
velocities show that they are only weak in the moss; 2.7 km/s (red-shifted) and 21.7 km/s, 
respectively, when averaged over time in all the 25 locations. Given the uncertainties in
the absolute values, the Doppler velocity 
is consistent with a measurement of zero.
The average values of the non-thermal velocities in each of the 25 locations range from 17--31 km/s. 
The upper limit of this range is an extreme case: in all but this location the values are less than
26 km/s. These 
values are no more than the 
typical values measured for non-thermal mass motions in the quiet Sun at this temperature.
We confirmed this by performing double Gaussian fits to the \ion{Fe}{12} 195.119\,\AA\, line
in 39 quiet Sun synoptic datasets that we had processed for a differential emission measure
analysis \citep{brooks_etal2009}. The data were taken with the EIS 1$''$ slit for the best
spectral resolution, and the non-thermal velocity we obtained for the quiet Sun was 25 km/s.
The standard deviation in the moss measurements is less than $\sim$ 15\% of the 
average values for each location over the 16 hours of observation, which corresponds to a variation of less than $\sim$ 4 km/s
in all the boxed areas. 
The average Doppler velocities are also less than $\sim$ 6.5 km/s in each of 
the regions. This is consistent with the measurements of only a few km/s in
the moss by \citet{warren_etal2008b}. 
The Doppler velocities also
vary by less than 3 km/s during the observations period. As with the intensity results, this is
less than the measured 
variation from location to location.

\section{Conclusions} 
Active region 10960 evolved slowly 
when seen in XRT images, and the moss in the core of the region also developed
slowly in TRACE and EIS images. The region was flare productive, however, with many
C- and M- class flares occurring throughout its passage across the
solar disk. Several of these flares occurred close to the neutral line in the core of the active region
during the 16 hours of data we analyzed with the potential to interact
with and disrupt the moss. AR 10960 was
therefore a good candidate for studying the relative importance of impulsive and steady
heating in the core of an active region. 

The data we presented are the highest cadence spatially resolved maps of non-thermal and Doppler
velocities in moss ever obtained in the corona. As such they are ideal for detecting any signatures
of dynamics in the moss, should they exist. Our spectrally resolved measurements of 
\ion{Fe}{12} 195.119\,\AA\, intensities agree with previous results from broad-band filter
images in that the intensities do not vary significantly in the moss. 
Our new measurements
of Doppler and non-thermal velocities show that they are only weak in the moss, and 
that they vary by only a few km/s during the observing period. 

The results confirm the findings of \citet{antiochos_etal2003} that moss at the bases of high temperature coronal loops is 
heated quasi-steadily `to an excellent approximation'. Perhaps surprisingly, we find that this is also true 
in the core of a flare productive active region, where the magnetic field is clearly changing.
It may be that steady heating dominates in the moss in active region cores,
with the variability being due to obscuration by overlying spicular material
as suggested by \citet{depontieu_etal1999}. In any case, it is clear that
the heating does not give rise to strong flows or motions in the moss, and that
the time-scale between impulsive heating events has to be short enough to maintain loops
at high temperatures, in contrast to models that assume loops have time to cool substantially
between events. 

\acknowledgements
We thank Ignacio Ugarte-Urra and the referee for helpful suggestions.
{\it Hinode} is a Japanese mission developed and launched by ISAS/JAXA,
with NAOJ as domestic partner and NASA and STFC (UK) as international partners.
It is operated by these agencies in co-operation with ESA and NSC (Norway).


\begin{thebibliography}{}

\bibitem[\protect\citeauthoryear{{Antiochos} et~al.}{{Antiochos}
  et~al.}{2003}]{antiochos_etal2003}
{Antiochos}, S.~K., {Karpen}, J.~T., {DeLuca}, E.~E., {Golub}, L.,  \&
  {Hamilton}, P. 2003, \apj, 590, 547

\bibitem[\protect\citeauthoryear{{Berger} et~al.}{{Berger}
  et~al.}{1999a}]{berger_etal1999}
{Berger}, T.~E., {de Pontieu}, B., {Fletcher}, L., {Schrijver}, C.~J.,
  {Tarbell}, T.~D.,  \& {Title}, A.~M. 1999a, \solphys, 190, 409

\bibitem[\protect\citeauthoryear{{Berger} et~al.}{{Berger}
  et~al.}{1999b}]{tberger_etal1999}
{Berger}, T.~E., {de Pontieu}, B., {Schrijver}, C.~J.,  \& {Title}, A.~M.
  1999b, \apjl, 519, L97

\bibitem[\protect\citeauthoryear{{Brooks}, {Ugarte-Urra}, \& {Warren}}{{Brooks}
  et~al.}{2008}]{brooks_etal2008}
{Brooks}, D.~H., {Ugarte-Urra}, I.,  \& {Warren}, H.~P. 2008, \apjl, 689, L77

\bibitem[\protect\citeauthoryear{{Brooks} \& {Warren}}{{Brooks} \&
  {Warren}}{2008}]{brooks&warren_2008}
{Brooks}, D.~H.,  \& {Warren}, H.~P. 2008, \apj, 687, 1363

\bibitem[\protect\citeauthoryear{{Brooks} et~al.}{{Brooks}
  et~al.}{2009}]{brooks_etal2009}
{Brooks}, D.~H., {Warren}, H.~P., {Williams}, D.~R.,  \& {Watanabe}, T. 2009,
  \apj

\bibitem[\protect\citeauthoryear{{Brown} et~al.}{{Brown}
  et~al.}{2008}]{brown_etal2008}
{Brown}, C.~M., {Feldman}, U., {Seely}, J.~F., {Korendyke}, C.~M.,  \& {Hara},
  H. 2008, \apjs, 176, 511

\bibitem[\protect\citeauthoryear{{Culhane} et~al.}{{Culhane}
  et~al.}{2007}]{culhane_etal2007}
{Culhane}, J.~L., et~al. 2007, \solphys, 243, 19

\bibitem[\protect\citeauthoryear{{de Pontieu} et~al.}{{de Pontieu}
  et~al.}{1999}]{depontieu_etal1999}
{de Pontieu}, B., {Berger}, T.~E., {Schrijver}, C.~J.,  \& {Title}, A.~M. 1999,
  \solphys, 190, 419

\bibitem[\protect\citeauthoryear{{de Pontieu}, {Tarbell}, \& {Erd{\'e}lyi}}{{De
  Pontieu} et~al.}{2003}]{depontieu_etal2003}
{de Pontieu}, B., {Tarbell}, T.,  \& {Erd{\'e}lyi}, R. 2003, \apj, 590, 502

\bibitem[\protect\citeauthoryear{{Doschek} et~al.}{{Doschek}
  et~al.}{2008}]{doschek_etal2008}
{Doschek}, G.~A., {Warren}, H.~P., {Mariska}, J.~T., {Muglach}, K., {Culhane},
  J.~L., {Hara}, H.,  \& {Watanabe}, T. 2008, \apj, 686, 1362

\bibitem[\protect\citeauthoryear{{Golub} et~al.}{{Golub}
  et~al.}{2007}]{golub_etal2007}
{Golub}, L., et~al. 2007, \solphys, 243, 63

\bibitem[\protect\citeauthoryear{{Handy} et~al.}{{Handy}
  et~al.}{1999}]{handy_etal1999}
{Handy}, B.~N., et~al. 1999, \solphys, 187, 229

\bibitem[\protect\citeauthoryear{{Harra} et~al.}{{Harra}
  et~al.}{2008}]{harra_etal2008}
{Harra}, L.~K., {Sakao}, T., {Mandrini}, C.~H., {Hara}, H., {Imada}, S.,
  {Young}, P.~R., {van Driel-Gesztelyi}, L.,  \& {Baker}, D. 2008, \apjl, 676,
  L147

\bibitem[\protect\citeauthoryear{{Klimchuk}}{{Klimchuk}}{2006}]{klimchuk_2006}
{Klimchuk}, J.~A. 2006, \solphys, 234, 41

\bibitem[\protect\citeauthoryear{{Kosugi} et~al.}{{Kosugi}
  et~al.}{2007}]{kosugi_etal2007}
{Kosugi}, T., et~al. 2007, \solphys, 243, 3

\bibitem[\protect\citeauthoryear{{Martens}, {Kankelborg}, \&
  {Berger}}{{Martens} et~al.}{2000}]{martens_etal2000}
{Martens}, P.~C.~H., {Kankelborg}, C.~C.,  \& {Berger}, T.~E. 2000, \apj, 537,
  471

\bibitem[\protect\citeauthoryear{{Patsourakos} \& {Klimchuk}}{{Patsourakos} \&
  {Klimchuk}}{2008}]{patsourakos&klimchuk_2008}
{Patsourakos}, S.,  \& {Klimchuk}, J.~A. 2008, \apj, 689, 1406

\bibitem[\protect\citeauthoryear{{Sakao} et~al.}{{Sakao}
  et~al.}{2007}]{sakao_etal2007}
{Sakao}, T., et~al. 2007, Science, 318, 1585

\bibitem[\protect\citeauthoryear{{Shimizu} et~al.}{{Shimizu}
  et~al.}{2007}]{shimizu_etal2007}
{Shimizu}, T., et~al. 2007, \pasj, 59, 845

\bibitem[\protect\citeauthoryear{{Tripathi} et~al.}{{Tripathi}
  et~al.}{2008}]{tripathi_etal2008}
{Tripathi}, D., {Mason}, H.~E., {Young}, P.~R.,  \& {Del Zanna}, G. 2008, \aap,
  481, L53

\bibitem[\protect\citeauthoryear{{Ugarte-Urra}, {Warren}, \&
  {Brooks}}{{Ugarte-Urra} et~al.}{2009}]{ugarteurra_etal2009}
{Ugarte-Urra}, I., {Warren}, H.~P.,  \& {Brooks}, D.~H. 2009, \apj, 695, 642

\bibitem[\protect\citeauthoryear{{Warren} \& {Winebarger}}{{Warren} \&
  {Winebarger}}{2006}]{warren&winebarger_2006}
{Warren}, H.~P.,  \& {Winebarger}, A.~R. 2006, \apj, 645, 711

\bibitem[\protect\citeauthoryear{{Warren} \& {Winebarger}}{{Warren} \&
  {Winebarger}}{2007}]{warren&winebarger_2007}
{Warren}, H.~P.,  \& {Winebarger}, A.~R. 2007, \apj, 666, 1245

\bibitem[\protect\citeauthoryear{{Warren} et~al.}{{Warren}
  et~al.}{2008}]{warren_etal2008b}
{Warren}, H.~P., {Winebarger}, A.~R., {Mariska}, J.~T., {Doschek}, G.~A.,  \&
  {Hara}, H. 2008, \apj, 677, 1395

\bibitem[\protect\citeauthoryear{{Young} et~al.}{{Young}
  et~al.}{2009}]{young_etal2009}
{Young}, P.~R., {Watanabe}, T., {Hara}, H.,  \& {Mariska}, J.~T. 2009, \aap,
  495, 587

\end{thebibliography}

\end{document}